\documentstyle[12pt]{article}
\begin{document}

\begin{titlepage}
\rightline{\large February 2007}

\vskip 2cm

\centerline{\Large \bf A simple explanation of the PVLAS anomaly }

\vskip 0.2cm
\centerline{\Large \bf in spontaneously 
broken mirror models}

\vskip 3cm
\centerline{R. Foot and A. Kobakhidze\footnote{E-mail: foot@physics.unimelb.edu.au, archilk@physics.unimelb.edu.au}}
\vskip 1cm
\centerline{\it School of Physics} 
\vskip 0.2cm
\centerline{\it Research Centre for High Energy Physics}
\vskip 0.2cm
\centerline{University of Melbourne,}
\vskip 0.2cm
\centerline{\it Victoria 3010 Australia}
\vskip 1cm

\noindent 
The PVLAS anomaly can be explained if there exist millicharged particles of 
mass $\stackrel{<}{\sim} 0.1$ eV and electric charge $\epsilon \sim 10^{-6} e$. We point out that such
 particles occur naturally in spontaneously broken mirror models. We argue 
that this interpretation of the PVLAS anomaly is not in conflict with astrophysical
 constraints 
due to the self interactions of the millicharged particles 
which lead them to be trapped within stars. This conclusion also holds for a generic paraphoton model.

\end{titlepage}

The PVLAS collaboration has obtained evidence for an anomalously large 
rotation of the polarization plane of light after its passage through a 
transverse magnetic field
 in vacuum\cite{pvlas}. One possible explanation\cite{gies,gies2} of this result 
requires there to exist millicharged
 particles of mass 

\begin{eqnarray}
m_f \stackrel{<}{\sim} 0.1 \ eV
\end{eqnarray} and electric charge

\begin{eqnarray}
\epsilon \sim 10^{-6} e \ .
\end{eqnarray}
Such millicharged particles can arise naturally in gauge models
 with a hidden sector containing an unbroken $U(1)$ gauge factor
 with a kinetic mixing term with the standard $U(1)_Y$ gauge 
field\cite{holdom,flv,fl,flv2}.

We will show in this letter that a gauge model of this general type can explain the PVLAS
experiment whilst being consistent with astrophysical constraints. However, rather than 
work in the framework of a generic hidden sector model we consider
the specific case of a spontaneously broken mirror model because such a model can also
simply explain the required tiny masses for the millicharged particles.

In the mirror model the hidden sector is exactly the same as the standard 
sector, so that the gauge group is $G_{SM} \otimes
 G_{SM} (G_{SM} \equiv SU(3)_c \otimes SU(2)_L \otimes U(1)_Y$)\cite{flv,fl,flv2}.
For each ordinary particle there is a mirror partner, which we denote with a prime ($'$). An exact Lagrangian $Z_2$ symmetry  
interchanging ordinary and mirror particles is 
hypothesised, which means that all the couplings
in the mirror sector are the same as in the ordinary sector. Note that the $Z_2$ symmetry can be interpreted as a parity symmetry ($x \to -x$) if the roles of left and right chiral fermion fields are interchanged in the mirror sector.

It is known\cite{flv} that there are only 2 renormalizable and gauge invariant
 Lagrangian terms coupling the ordinary and mirror sector together: $U(1)_Y$ - $U(1)'_Y$ 
gauge boson kinetic mixing, and ordinary-mirror Higgs scalar interactions, i.e. 
\begin{eqnarray}
{\epsilon \over 2}F_{\mu \nu}' F^{\mu \nu}
\end{eqnarray}
and 
\begin{eqnarray}
\lambda \phi'^{\dagger}\phi' \phi^{\dagger} \phi.
\end{eqnarray}
Most work on this model has focussed on the case where the $Z_2$ 
symmetry is unbroken by the vacuum, $\langle \phi \rangle = \langle
\phi' \rangle$, which means that the mirror particles 
have the same masses as their ordinary counterparts\cite{flv}.
However, there is another vacuum solution to 
the Higgs potential 
where the $Z_2$ symmetry is spontaneously broken,
with $\langle \phi \rangle = u \simeq 174$ GeV and $\langle \phi' \rangle = 0$ \cite{fl,flv2}.
This vacuum solution can be made manifest by parameterizing the 
most general Higgs potential in the form:
\begin{eqnarray}
V(\phi, \phi') = \lambda_1 \left( \phi^{\dagger} \phi + \phi'^{\dagger}\phi'
- u^2\right)^2
\ + \ 
\lambda_2 \phi'^{\dagger}\phi' \phi^{\dagger}\phi \ . 
\end{eqnarray}
 Written in this way, $V \ge 0$ for $\lambda_1, \lambda_2 > 0$,
and $V = 0$ iff $\langle \phi \rangle = u, \ \langle \phi' \rangle = 0$ 
(or $\langle \phi' \rangle = u, \ \langle \phi \rangle = 0$).

When dynamical effects from mirror QCD condensation are 
included $\langle \phi' \rangle \neq 0$.
This effect arises from the Yukawa coupling terms:
$h_q \bar q'_R q'_L \phi' + H.c.$, which induce a linear term in $\phi'$ in
 the Higgs potential [given 
that the mirror quarks condense: $\langle \bar q' q' \rangle =
\Lambda'^3$, with $\Lambda' \sim 100$ MeV, see Ref.\cite{fl, flv2} for more
 details]. The net effect is to induce a small but non-zero VEV for
$\langle \phi' \rangle$:
\begin{eqnarray} 
\langle \phi' \rangle \simeq h_t \Lambda'^3/m_{\phi'_0}^2 
\end{eqnarray} where $h_t \approx 1$ is the mirror top-quark Yukawa coupling. 
It follows that these spontaneously broken mirror models have 
their mirror fermion masses suppressed by a factor $\eta \equiv 
\langle \phi'\rangle/\langle \phi \rangle$, with 
\begin{eqnarray}
\eta & \simeq & {h_t \Lambda'^3 \over u m^2_{\phi'_0} }
\nonumber \\
& \sim & 10^{-7} \left( {10 \ {\rm GeV} \over m_{\phi'_0} }\right)^2
 \ . \end{eqnarray} Thus the masses of the mirror fermions are naturally in the sub eV 
range - a necessary requirement to explain the PVLAS anomaly.
 Note that the mass of the mirror scalar is $m^2_{\phi'_0} \simeq \lambda_2 u^2$, and is experimentally weakly constrained
because the scalar only couples very weakly to the ordinary particles via cubic and quartic
interactions with the standard Higgs scalar ($\phi$)  
\footnote{It is possible to decouple $\eta$ from $m_{\phi'_0}$, by adding a
$Z_2$ singlet scalar to the theory, which allows
the VEV of $\phi'$ to be made into a completely free parameter\cite{mohap}.}\cite{flv2}.

It is generally argued that the astrophysical limits on epsilon 
are very stringent, $\epsilon \stackrel{<}{\sim} 10^{-14}$, arising from 
energy loss considerations in stars\cite{raf1,raf2}.
Interestingly though, the published limits apply only in the millicharged mass region above about 1 eV
(see e.g. figure 1 of ref.\cite{raf2}). For masses significantly above around 1 eV,
the number density of millicharged particles in stars can become Boltzman suppressed
($exp(-m_{e'}/T)$). In this situation, the mirror photon path length becomes
very long and the photons can freely stream out of the hot inner core of the star - leading
to the stringent limit $\epsilon \stackrel{<}{\sim} 10^{-14}$ mentioned above.
It turns out that the case of light milli charged particles - with masses in
the sub eV range is quite different. Because the masses are less than the temperature
within the star, there is no Boltzman supression - and the mean free path of the mirror
photons becomes very short. That is, the mirror particles 
are trapped within the stars by interactions. Let us now examine the physics in more detail.

There are basically two type of interactions: firstly, the mirror
 particles can interact with the ordinary particles
 which will produce a thermal population of $\bar e', e', \gamma'$.
For example, $\bar e', e'$ pairs can be produced via 
virtual Bremsstrahlung photons: $p+e \to p+e + \bar e' + e'$.  
Secondly, there are also 
self interactions, such as $e'+\bar e'
\leftrightarrow \gamma' + \gamma'$, $e' \gamma' \to e' \gamma'$.
 The cross section for the Bremsstrahlung process, and elastic
scattering processes such as $e' e \leftrightarrow e'e$ are proportional to
$\epsilon^2$ and for $\epsilon \sim 10^{-6}$ are large enough to 
locally thermalize the mirror particles with the ordinary ones
throughout the entire region of the star. 

To see this, consider the $e'e \to e'e$ process as an example.
If $m_{e'} \stackrel{<}{\sim} 0.1$ eV, the light mirror
particles $\gamma', e', \bar e'$ form a relativistic gas.
The cross section for the $e' e \to e' e$ elastic scattering process is of order:
\begin{eqnarray}
\sigma \sim {\epsilon^2 \alpha^2 \over s}
\end{eqnarray}
so that the $e'-e$ scattering length, $d$, is:
\begin{eqnarray}
d &=& {1 \over \sigma n_e} \nonumber \\
 & \sim & {(T/eV)^2 \over (\epsilon/10^{-6})^2} \left( {g/cm^3 \over
\rho}\right) \ {\rm meters}
\end{eqnarray}
Thus, in the sun, for example, $d$ ranges from about 10 kilometers
in the core to of order a meter in the outer regions. This distance
is so short that the $e'$ should be in local thermodynamic equilibrium
with the ordinary matter in the sun, to a very good approximation.
This means that at each point
within the star a single temperature $T(r)$ may be defined. In particular, 
the mirror particles cannot have any net outward
velocity.

Let us now consider the self interactions of the mirror particles.
The cross section for the self interactions of the mirror particles 
are not suppressed by 
$\epsilon$ and can be quite large because the mirror particles $e',\bar
 e'$ are so light.
The cross section for the  
self interations of the mirror particles,
for processes 
such as $e' + \bar e' \leftrightarrow \gamma' + \gamma'$, is thus relatively 
large, being of order:
\begin{eqnarray}
\sigma \sim {\alpha^2 \over s}  
\end{eqnarray}
leading to an interaction length of 
\begin{eqnarray}
{\ell'} &=& \langle {1 \over \sigma n} \rangle \nonumber \\
&\sim & {1 \over \alpha^2 T} \
\sim  {10^{-4} \ {\rm cm} \over T/keV } \ . 
\label{xy}
\end{eqnarray}
The small mean free path of the mirror particles can alleviate 
their impact on the transport of energy within stars - which is another
 important astrophysical constraint on exotic particles weakly 
coupled to the ordinary particles.
Typically,  
exotic particles shouldn't 
transport energy faster than energy transport from ordinary photons\cite{raf1}.

The relevant equation governing the radiative energy transport is:
\begin{eqnarray}
L_r = - {4\pi r^2 \over 3\kappa_{\gamma} \rho} {d(aT^4) \over dr}
\label{rad}
\end{eqnarray} 
where $L_r$ is the
interior luminosity due to all of the energy generated within the star interior to the radius $r$. Also 
$aT^4$ is the energy density of the radiation field ($a =
\pi^2/15$ in natural units) and $\kappa_{\gamma}$ is the opacity. By definition, 
$(\kappa_{\gamma} \rho)^{-1} \equiv \ell_{\gamma}$ is the photon mean free path.
Thus, as the photon mean free path becomes smaller, energy is
 transported less efficiently (for a given temperature gradient). Physically this 
is because radiative energy transport is essentially a random walk process.
If we include the contribution of the mirror particles, and write the 
energy transport equation in terms of the mean free paths, then we have:
\begin{eqnarray}
L_r = - {4\pi r^2 \over 3} \left(\ell_{\gamma} + \ell_{\gamma'} + \ell_{e'} +
\ell_{\bar e'}\right) {d(aT^4) \over dr}\ . 
\label{rad2}
\end{eqnarray}
Note that $\ell_{\gamma'}, \ell_{e'}, \ell_{\bar e'} \sim \ell'$
 (Eq.\ref{xy}).
 Thus, the mirror particles do not contribute significantly to
 energy transport
 if $\ell' \stackrel{<}{\sim} \ell_{\gamma}$.
The photon mean free path is:
\begin{eqnarray}
\ell_{\gamma} &=& {1 \over \kappa_{\gamma} \rho}
\end{eqnarray} with $\kappa_{\gamma}$ typically of order $\kappa^{-1}_{\gamma} \sim 1 \ {\rm g/cm^3}$.  For the conditions within the interior of main sequence stars, $\rho
\stackrel{<}{\sim} 10^2 \ {\rm g/cm^3}$ and hence ${\ell'} \ll
\ell_{\gamma}$. 
Evidently, the mirror particles are typically consistent 
with the energy transfer constraints. 
We therefore conclude that this type of explanation for the PVLAS anomaly appears to be consistent 
with astrophysical constraints. This conclusion should also hold for a generic hidden sector model 
(sometimes called `paraphoton' models in the literature) provided that the 
model has qualitatively similar features to the spontaneously broken mirror model. That is, has the milli electric charges of the hidden sector fermions or 
bosons\footnote{
The experimental implications of millicharged bosons are quite different
to fermions\cite{gies2}.}
induced by the kinetic mixing of the photon with a massless (or very light) hidden sector gauge boson.

Of course, even if the model is 
consistent with astrophysical bounds, the model will 
still be problematic for Big Bang Nucleosynthesis (BBN). This is because there will be a
 significant increase in energy density arising from the production
 of mirror particles, which will typically thermalize with the ordinary
 ones at the BBN epoch ($T \sim$ 1 MeV). However, 
the increase in energy density, could be compensated, by e.g. an
 electron neutrino asymmetry\cite{neu}. Therefore, the millicharged particle 
interpretation of the PVLAS anomaly cannot be rigorously excluded via BBN arguments. Importantly, the millicharged particle interpretation will be experimentally tested in the near future by
forthcoming experiments (see. e.g. discussion in Ref.\cite{gies2,gen}).

Another important issue is the impact of the model on the cosmic microwave background (CBM)
measurements.
The existence of the light mirror particles in local thermodynamic
equilibrium with
the photons might be expected to have some important effects at the
recombination era because the pressure and density of the
relativistic component is increased relative to the baryonic part (c.f.
ref.\cite{mel}).
To determine the significance of these effects
a careful study might therefore be needed 
varying all of the known parameters (including $\Omega_b$). 
Post recombination, observe that
the photons remain in thermal equilibrium with these light mirror particles,
via processes such as $\gamma' + \gamma \leftrightarrow e' + \bar e'$. The
cross section for these processes is relatively large 
(for $T \stackrel{>}{\sim} m_{e'}$): $\sigma \sim \epsilon^2 \alpha^2/T^2$, leading to an interaction rate of order:
\begin{eqnarray}
\Gamma &\sim &\epsilon^2 \alpha^2 T \nonumber \\
& \sim & 0.1 \left({\epsilon \over 10^{-6}}\right)^2 \left( {T \over eV} \right)\ s^{-1}
\end{eqnarray}
Note that this interaction rate is very many orders of magnitude larger than the explansion rate of the Universe, so that the interactions will not lead to any spectral distortions
in the CMB.
When $T \approx m_{e'}$, the $e'$ will begin to annhilate and heat the
remaining $\gamma, \gamma'$. The main effect of this heating of the CMB should simply
be to increase the inferred lifetime of the Universe. Since without the heating effect,
the current CMB temperature should be lower than what it is measured to be - and this would happen if the Universe were older so that a larger red shift could occur.

In conclusion, we have pointed out that the PVLAS anomaly might be explained within the context of spontaneously broken mirror models (or a generic variation, such as a generic
`paraphoton' model). These models contain a spectrum of light millicharged particles in the sub eV mass range. We have pointed out that such models
are {\it not} in conflict with
 astrophysical constraints due to the interactions of the millicharged particles
 which lead them to be trapped within stars (with a mean free path short enough to evade
constraints from energy transport within stars). 
However, the model is inconsistent with standard BBN, so if the millicharged particle interpretation of the PVLAS anomaly is experimentally confirmed, then a non-standard BBN scenario might have to be contemplated.

\vskip 0.5cm

\noindent
 {\bf Acknowledgements}: 
The authors would like to thank N. Cornish and J. Redondo for useful correspondence. 
This work was supported by the Australian Research Council.


\begin{thebibliography}{999}


\bibitem{pvlas}
E. Zavattini {\it et al} (PVLAS Collaboration),
Phys. Rev. Lett. 96, 110406 (2006).





\bibitem{gies}
H. Gies, J. Jaeckel and A. Ringwald, Phys. Rev. Lett. 97, 140402 (2006) [
hep-ph/0607118].





\bibitem{gies2} M. Ahlers, H. Gies, J. Jaeckel and A. Ringwald, hep-ph/0612098.



\bibitem{holdom}
B. Holdom, Phys. Lett. B166, 196 (1986), see also L. B. Okun, Sov. Phys. JETP 56, 502 (1982).





\bibitem{flv}
R. Foot, H. Lew and R. R. Volkas, 
Phys. Lett. B272, 67 (1991); The concept of mirror matter was discussed prior to the advent of the standard model in T. D. Lee and C. N. Yang, Phys. Rev. 104, 256 (1956); I. Kobzarev, L. Okun and I. Pomeranchuk, Sov. J. Nucl. Phys. 3, 837 (1966); M. Pavsic, Int. J. Theor. Phys. 9, 229 (1974).





\bibitem{fl}
R. Foot and H. Lew, hep-ph/9411390 (1994).





\bibitem{flv2}
R. Foot, H. Lew and R. R. Volkas,
JHEP 0007, 032 (2000) [hep-ph/0006027].






\bibitem{mohap}
Z. Berezhiani and R. N. Mohapatra, 
Phys. Rev. D52, 6607 (1995) [hep-ph/9505385].





\bibitem{raf1}
G. G. Raffelt, {\it Stars as Laboratories for Fundamental Physics},
University of Chicago Press, 1996.





\bibitem{raf2}
S. Davidson, S. Hannestad and G. Raffelt,
JHEP 0005, 003 (2000) [hep-ph/0001179].






\bibitem{neu}
See e.g. K. A. Olive, D. N. Schramm, D. Thomas and T. P. Walker, Phys. Lett. B265, 239 (1991).



\bibitem{gen}
S. N. Gninenko, N. V. Krasnikov and A. Rubbia, hep-ph/0612203.

\bibitem{mel}
A. Melchiorri, A. Polosa and A. Strumia, hep-ph/0703144.

\end{thebibliography}
\end{document}